\documentclass[12pt,a4paper]{article}
\usepackage[dvips]{graphicx} 

\begin{document}
\begin{center}
{\Large \bf Monte Carlo Simulation
of a Random-Field Ising Antiferromagnet}
\vspace{0.5cm}

{\bf V. V. Prudnikov and V. N. Borodikhin\\
Omsk State University, pr. Mira 55, Omsk, 644077 Russia\\
e-mail: prudnikv@univer.omsk.su}
\end{center}

\begin{abstract}
Phase transitions in the three-dimensional diluted Ising antiferromagnet in an applied magnetic
field are analyzed numerically. It is found that random magnetic field in a system with spin concentration below
a certain threshold induces a crossover from second-order phase transition to first-order transition to a new
phase characterized by a spin-glass ground state and metastable energy states at finite temperatures.
\end{abstract}

\section{Introduction}

Critical behavior of disordered systems with
quenched disorder has been the subject of much theoretical
and experimental interest, because the presence
of quenched defects in most real solids modifies their
thermodynamic characteristics, including critical
behavior. It is well known that quenched disorder manifests
itself by temperature fluctuations in ferro- and
antiferromagnetic systems in the absence of external
magnetic field or by magnetic field fluctuations in antiferromagnets
in uniform magnetic field.

In the former case, quenched disorder affects the
properties of only those homogeneous magnetic
materials whose specific heat is divergent at the critical
point \cite{1}. Otherwise, the presence of defects does not
change the critical behavior of magnets. This criterion
applies only when the effective Hamiltonian near the
critical point is isomorphic to the Ising model Hamiltonian.
Disorder-induced critical behavior of the Ising
model was analyzed in numerous recent studies \cite{2}. For
dilute Ising-like systems, it was found that theoretical
calculations are in good agreement with experimental
results and Monte Carlo simulations.

Despite extensive theoretical and experimental studies
of random-field magnets conducted over the past
twenty years \cite{3}, very few facts concerning their behavior
have been established. In particular, the nature of
phase transition in the random-field Ising model
remains unclear, and the currently available theoretical
results in this area disagree with experiment. The only
theoretically proved fact is that the upper critical
dimension for this phase transition is six (i.e., critical
phenomena in systems of higher dimension are
described by mean field theory) \cite{3}, whereas the critical
dimension is four for homogeneous systems. While it
had been argued that the lower critical dimension $d_l$
can be both $d_l=2$ \cite{4} and $d_l=3$ \cite{5}
(i.e., there is long-range
order at finite temperatures if the system's dimension is
higher), specialists came to the conclusion that $d_l=2$
after the publication of \cite{6, 7}. However, the nature of
phase transition in the three-dimensional random-field
Ising model remains unclear. According to \cite{8, 9}, it is a
first-order phase transition even at very low random-
field strengths; according to \cite{10, 11}, it is a secondorder
transition.

The effect of random fields on the behavior of magnetic
systems is described by using two qualitatively
equivalent models: the ferromagnetic random-field
Ising model (RFIM) \cite{12, 13} and the Ising diluted antiferromagnets
in a field (DAFF) \cite{14}. Real random-field
magnets are antiferromagnets with quenched nonmagnetic
impurities. Their behavior includes manifestations
of both antiferromagnetic interaction between
nearest neighbor atoms and ferromagnetic interaction
between next-nearest neighbor atoms. The structure of
an antiferromagnet can be represented as several interpenetrating
ferromagnetic sublattices such that the total
magnetization of the antiferromagnet is zero even
though each ferromagnetic sublattice is magnetically
ordered at a temperature below the Neel temperature.
Examples of two-sublattice antiferromagnets are the
following materials: $NiO$, $MnO$, $Fe_2O_3$, and $MnF_2$.
Examples of random-field magnets include the uniaxial
Ising-like antiferromagnets $MnF_2$ and $FeF_2$
diluted with zinc atoms in an external magnetic field \cite{15}.

\section{ Model }

In this study, a Monte Carlo method is used to simulate
the thermodynamic behavior of a diluted antiferromagnetic
Ising model in an applied magnetic field on
the simple cubic lattice by taking into account next-
nearest-neighbor interaction. The Hamiltonian of the
model has the form

\begin{equation}
  {\cal H}=J_{1}\sum_{i,j}p_ip_j\sigma_{i}\sigma_{j}+
  J_{2}\sum_{i,k}p_ip_k\sigma_{i}\sigma_{k}+\mu h\sum_{i}p_i\sigma_{i},
\end{equation}

where  $\sigma_{i}=\pm 1$ is the spin located at site i;
$\mu$ is the Bohr magneton; $J_{1}=1$ and $J_{2}=-1/2$
characterize antiferromagnetic nearest-neighbor and ferromagnetic nextnearest-
neighbor exchange couplings, respectively;
$h$ is the strength of the uniform magnetic field; and
$p_i$ and $p_j$ are random variables characterized by the distribution
function

\begin{equation}
P(p_i)=p\delta(p_i-1)+(1-p)\delta(p_i)
\end{equation}

which are introduced to describe quenched nonmagnetic
impurity atoms vacancies distributed over the lattice
and characterized by the concentration
$c_{imp}= 1-p$, where $p$ is spin concentration. For
$p=1.0$, the model
with competing interactions has been studied by Monte
Carlo methods for over twenty years \cite{16, 17}. The first
study of effects of disorder on critical behavior based
on this model was presented in \cite{18}. For the DAFF
mentioned above \cite{13}, competition between ferromagnetic
order parameters was not taken into account. This
model provides the most realistic physical representation.
Since the strength of random-field effects is determined
by impurity concentration and external field
strength both in the model and in real magnets, the
parameters of the model can be compared to those of
real physical experiments on Ising diluted antiferromagnets.
However, an analogous comparison of the
random field with the impurity concentration in a sample
and the applied field strength is difficult to perform
for the ferromagnetic random-field Ising model
(RFIM), which is most widely used in numerical simulations.
Therefore, random field variation in RFIM cannot
be quantitatively compared with structural disorder
in real systems, which is shown here to be the key factor
that controls phase transitions.

An antiferromagnet is characterized by the staggered
magnetization $M_{stg}$
defined as the difference of
the magnetizations of the two sublattices, which plays
the role of an order parameter. To determine the type of
phase transition, we calculate the Binder cumulant \cite{19}

\begin{equation}
U = \frac{1}{2}( 3 - \frac{[<M_{stg}^4>]}{[<M_{stg}^2>]^2}).
\end{equation}

where angle and square brackets denote statistical averaging
and averaging over disorder realizations. The calculation
of the cumulant is a good test for the order of
phase transition: the cumulants plotted versus tempera-
ture have a distinct point of intersection in the case of
second-order transition, whereas those corresponding
to first-order phase transition have a characteristic
shape and do not intersect.

We also examine spin-glass states. It is well known
that spin glasses are characterized by transition to a
phase with an infinite number of metastable states
separated by potential barriers in the thermodynamic
limit \cite{20}. The complex magnetic ordering in such systems
can be described in terms of the spin-glass order parameter

\begin{equation}
q_s=\frac{1}{pL^3}[<\sigma_i^{\alpha}\sigma_i^{\beta}>]
\end{equation}

where $\alpha$ and $\beta$
refer to the spin configurations corresponding
to replicas of the simulated disordered system
characterized by equal temperatures, but different initial
disorder realizations.

To obtain correct values of thermodynamic characteristics
of critical behavior, both statistical averaging
and averaging over disorder realizations must be performed
only after the system has thermalized. Critical
behavior of disordered systems is characterized by
anomalously long relaxation times, which rapidly
increase with the size of the simulated system. To reach
equilibrium at near-critical temperatures and determine
the corresponding thermodynamic characteristics, the
system was quenched with a temperature step of
$\Delta T=0.1$ starting from a temperature at which no metastable
states had been obtained in any sweep. At each temperature
step, a relaxation regime was computed in
$5000$ steps and averaging was performed in $10000$ steps
by using the spin configuration obtained at the preceding
step as an initial condition. This procedure was executed
to obtain a stable equilibrium at each temperature
and avoid metastable states \cite{18}.

For each lattice size $L$, thermodynamic characteristics
were computed for constant $h$ and $p$
by ensembleaveraging
the results of five sweeps executed for different
initial spin configurations corresponding to a particular
disorder realization and then averaging over $10$ to
$20$ different disorder realizations.

\section{ Results }

We examined the temperature dependence of several thermodynamic characteristics of three
- dimensional Ising antiferromagnets in a wide range of impurity concentrations for
systems having a size varying from $L=8$ to $L=64$ in applied magnetic fields of a
strength between $h=1$ and $h=4$.

Our analysis revealed several intervals of $p$
corresponding to different behavior for each value of $h$.
Second-order transition between paramagnetic and
ferromagnetic phases is observed at $T_c(h, p)$
when $p_u<p<1$ \cite{18}, where $p_u$ is the vacancy percolation
threshold ($p_u=0.83$ for the present model).

\begin{figure}
\includegraphics[width=0.47\textwidth]{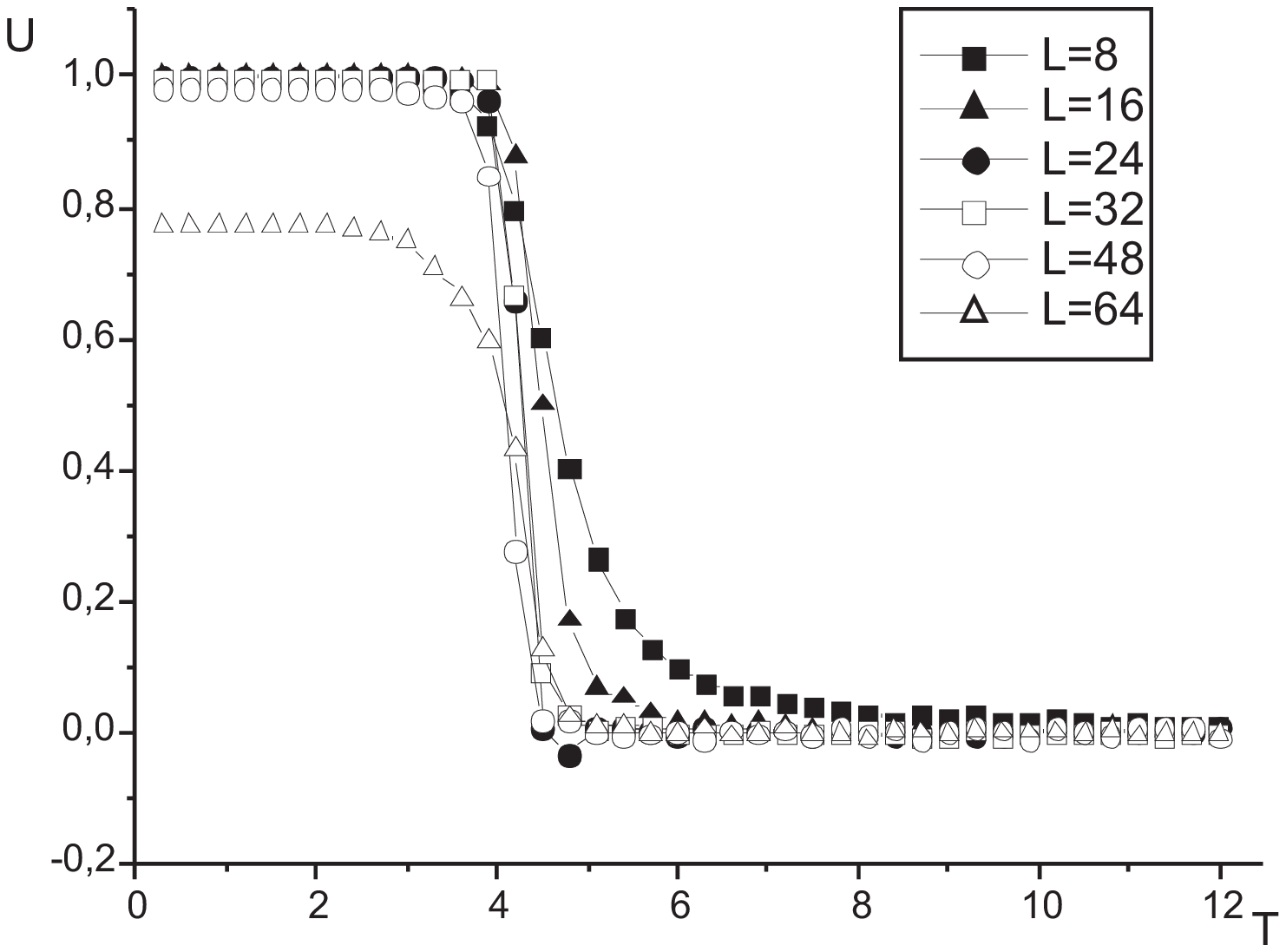} \hfill
\includegraphics[width=0.47\textwidth]{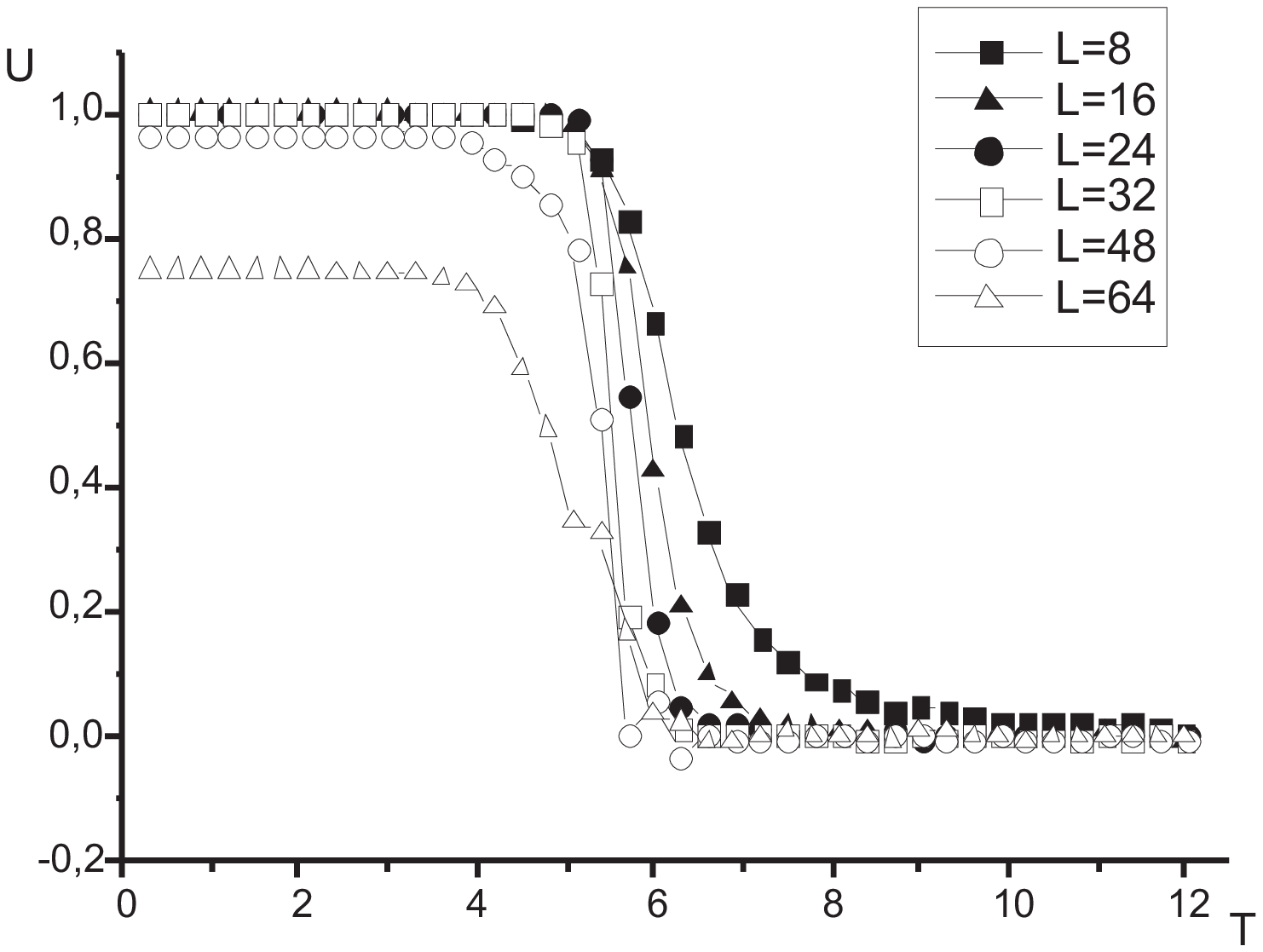} \\
\hspace*{0.23\textwidth} (a) \hfill (b) \hspace*{0.23\textwidth} \\
\includegraphics[width=0.47\textwidth]{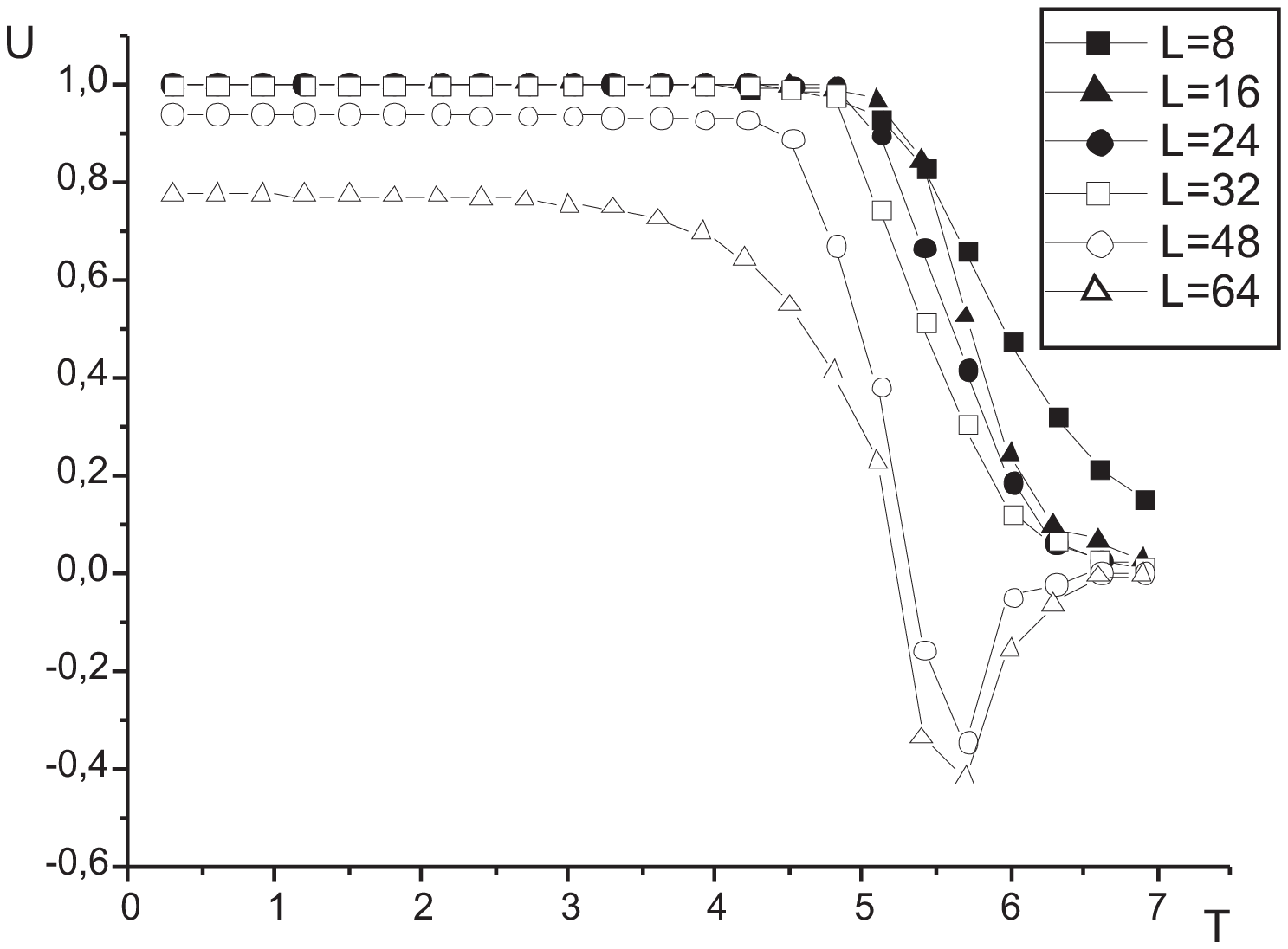} \hfill
\includegraphics[width=0.47\textwidth]{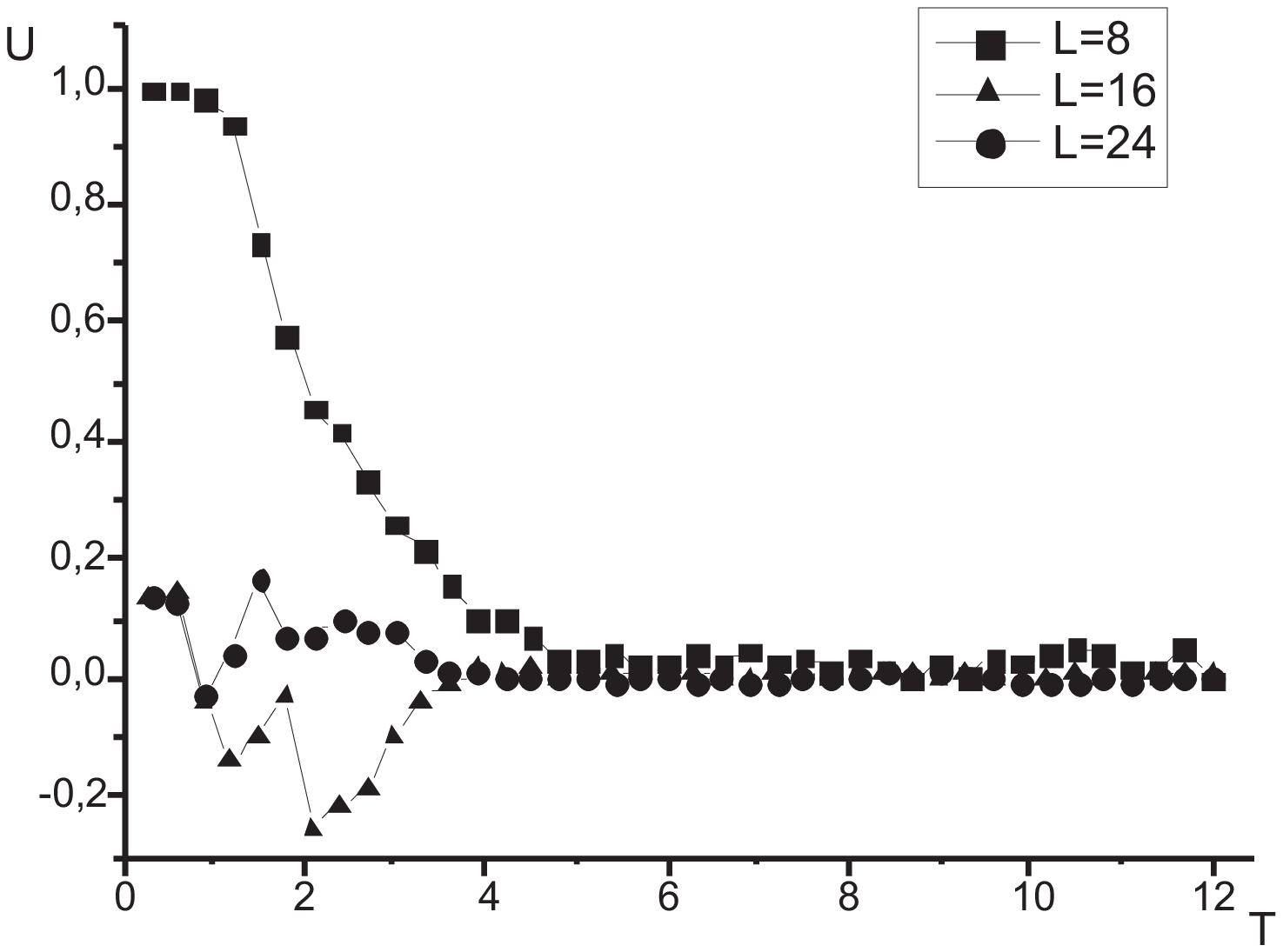} \\
\hspace*{0.23\textwidth} (c) \hfill (d) \hspace*{0.23\textwidth} \\
\caption{ Temperature dependence of the Binder cumulant on lattices with $L=8, 16 , 24,
32, 48$, and $64$: (a) $p=0.5$, $h=1$; (b) $p=0.725$, $h=3$; (c) $p=0.8$, $h=4$; (d)~$p=
0.5$, $h = 3$. }
\end{figure}

When $p_c<p<p_u$, where $p_c$ is the magnetic percolation threshold ( $p_c= 0.17$ for the
present model ), there exist such $p(L^{\prime}, h)$ that the computed quantities exhibit
behavior characteristic of sec ond- and first-order phase transition if $p>p(L^{\prime},
h)$ and $p<p(L^{\prime}, h)$, respectively, on lattices with $L<L^{\prime}$. The value of
$p(L^{\prime}, h)$ increases with $h$ and $L^{\prime}$, approaching the threshold $p_u =
0.83$.

This size-dependent behavior is explained by the
existence of interpenetrating spin and vacancy clusters
whose fractal dimensions vary between 0 and 3,
depending on spin concentration. Therefore, the sizedependent
parameterization of transition from longrange
order to domain structure with characteristic size
$L_c$ by

\begin{equation}
\frac{h_r}{J(L)}=\frac{h_r}{JL^{(2-d)/2}},  L_c\approx\left(\frac{J}{h_r}\right)^{2/(2-d)}
\end{equation}

proposed for Ising-like systems in \cite{21}, where $h_r$ is the
random-field amplitude, $J$ is the exchange coupling,
and $d_f$ is interpreted as the fractal dimension of the spin
cluster, can be used to predict that antiferromagnetic
long-range order breaks down at $d_f < 2$.

Figures 1-4 illustrate the existence of boundaries
separating spin-concentration intervals characterized
by different strength of random-field effects for systems
with $L\leq 64$ in applied magnetic fields of strength
between $h=1$ and $h=4$.

Figure 1 shows the temperature-dependent Binder
cumulants calculated for several lattices with $p=0.5$ for
$h=1$, with $p=0.5$ and $0.725$ for $h=3$, and with $p=0.8$
for $h=4$. For spin concentrations close to $p_u$, the Binder
cumulants do not intersect only if $L\geq 64$. When $p=0.5$
and $h=3$, no intersection of Binder cumulants is
observed for lattices of all sizes used in the computations.
Comparing Figs. 1a-1c, we see that the sizedependent
change in the behavior of Binder cumulants
due to the increase in field strength from $h=1$ to $h=4$
(increasing random-field effects) corresponds to the
spin concentration increasing from $p=0.5$ to $p=0.8$.
For systems with $p<p(L^{\prime}, h)$, the behavior of $M_{stg}(T)$
(Fig. 2) strongly depends on the lattice size for all values
 of $h$ used in the computations.

\begin{figure}
\includegraphics[width=0.47\textwidth]{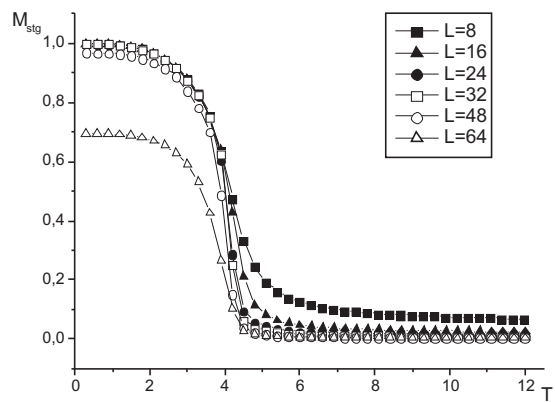} \hfill
\includegraphics[width=0.47\textwidth]{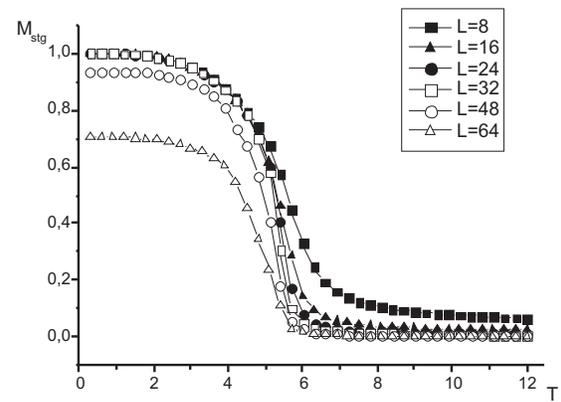} \\
\hspace*{0.23\textwidth} (a) \hfill (b) \hspace*{0.23\textwidth} \\
\includegraphics[width=0.47\textwidth]{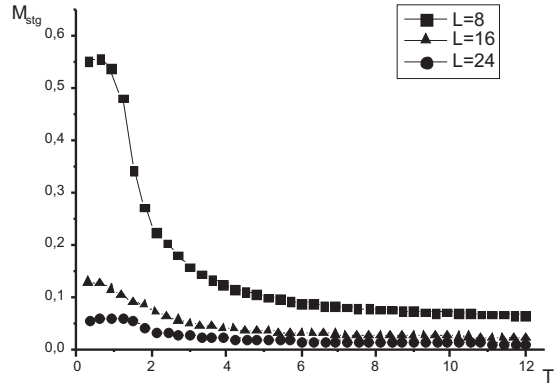} \\\vspace*{-0.5cm} 
\hspace*{0.23\textwidth}(c) \\\vspace*{-0.5cm}
\caption{ Temperature dependence of staggered magnetization on lattices with $L= 8, 16,
24, 32, 48,$ and $64$: (a) $p=0.5$, $h=1$; (b) $p=0.725$, $h=3$; (c) $p=0.5$, $h=3$. }
\end{figure}

\begin{figure}
\includegraphics[width=0.47\textwidth]{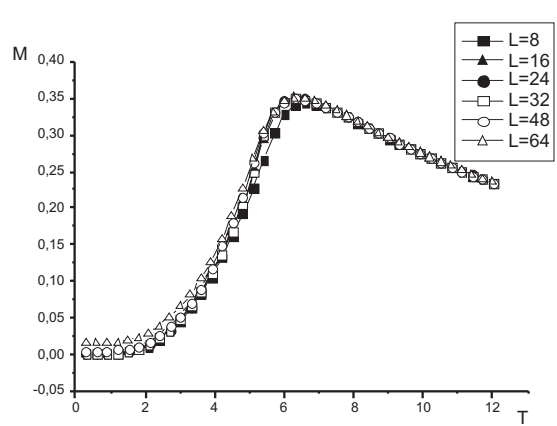} \hfill
\includegraphics[width=0.47\textwidth]{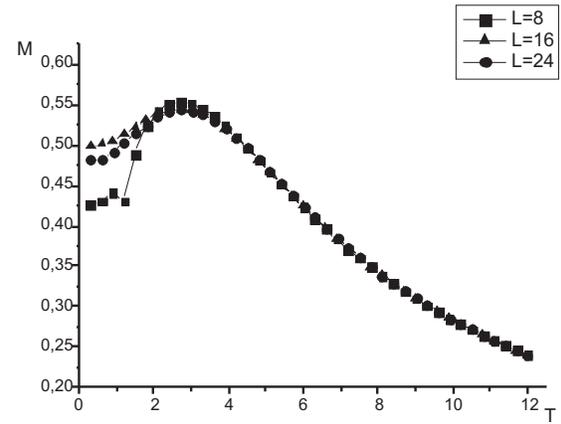} \\
\hspace*{0.23\textwidth} (a) \hfill (b) \hspace*{0.23\textwidth} \\
\caption{ Temperature dependence of total magnetization on lattices with $L=8, 16, 24,
32, 48$, and $64$: (a) $p=0.725$, $h=3$; (b) $p=0.5$, $h=3$. }
\end{figure}

The decrease in staggered
magnetization with increasing $L$ points to the
absence of an antiferromagnetic ground state. Furthermore,
the insignificant increase in total magnetization
$M$ with increasing $L$ (Fig. 3a) indicates that the system
breaks up into antiferromagnetic domains of size $L<L^{\prime}$
with nearly compensated magnetizations. As the random-
field effects increase with impurity concentration
and applied magnetic field, both number and size of
antiferromagnetic domains increases (Fig. 2c) and both
number and size of ferromagnetic ones increases
(Fig. 3b), while it holds that $M_{stg}+M<1$.

To further elucidate the properties of systems with
$p_c<p<p_u$, we examined the temperature dependence
of the spin-glass order parameter. The results obtained
for several disorder realizations are shown in Fig. 4.
The graphs demonstrate that a spin-glass phase with
"frozen" configuration of magnetic moments is
obtained as temperature approaches zero. Thus, a random
magnetic field induces transition from antiferromagnetic
to spin-glass ground state in the Ising model
with competing interactions when $p<p_u$. At finite temperatures,
the corresponding change in the state of a
disordered system is a first-order transition from a paramagnetic
to a mixed phase. When the spin concentration
is high, the latter consists of antiferromagnetic
domains separated by spin-glass regions. With decreasing
spin concentration, the number and size of antiferromagnetic
domains decrease and the number and size
of ferromagnetic domains increases, while the volume
fraction occupied by the spin-glass phase decreases.

\begin{figure}
\includegraphics[width=0.47\textwidth]{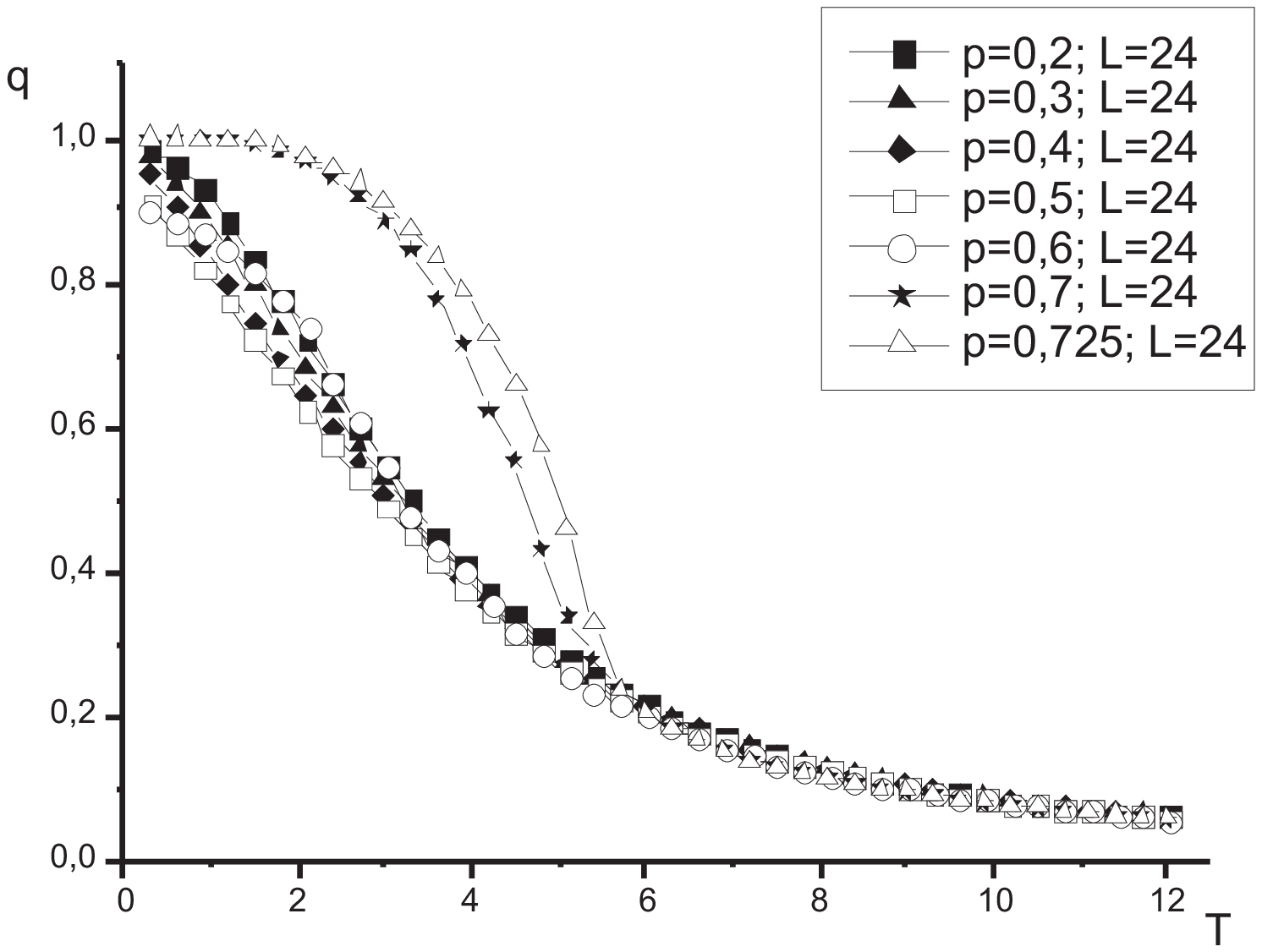} \hfill
\includegraphics[width=0.47\textwidth]{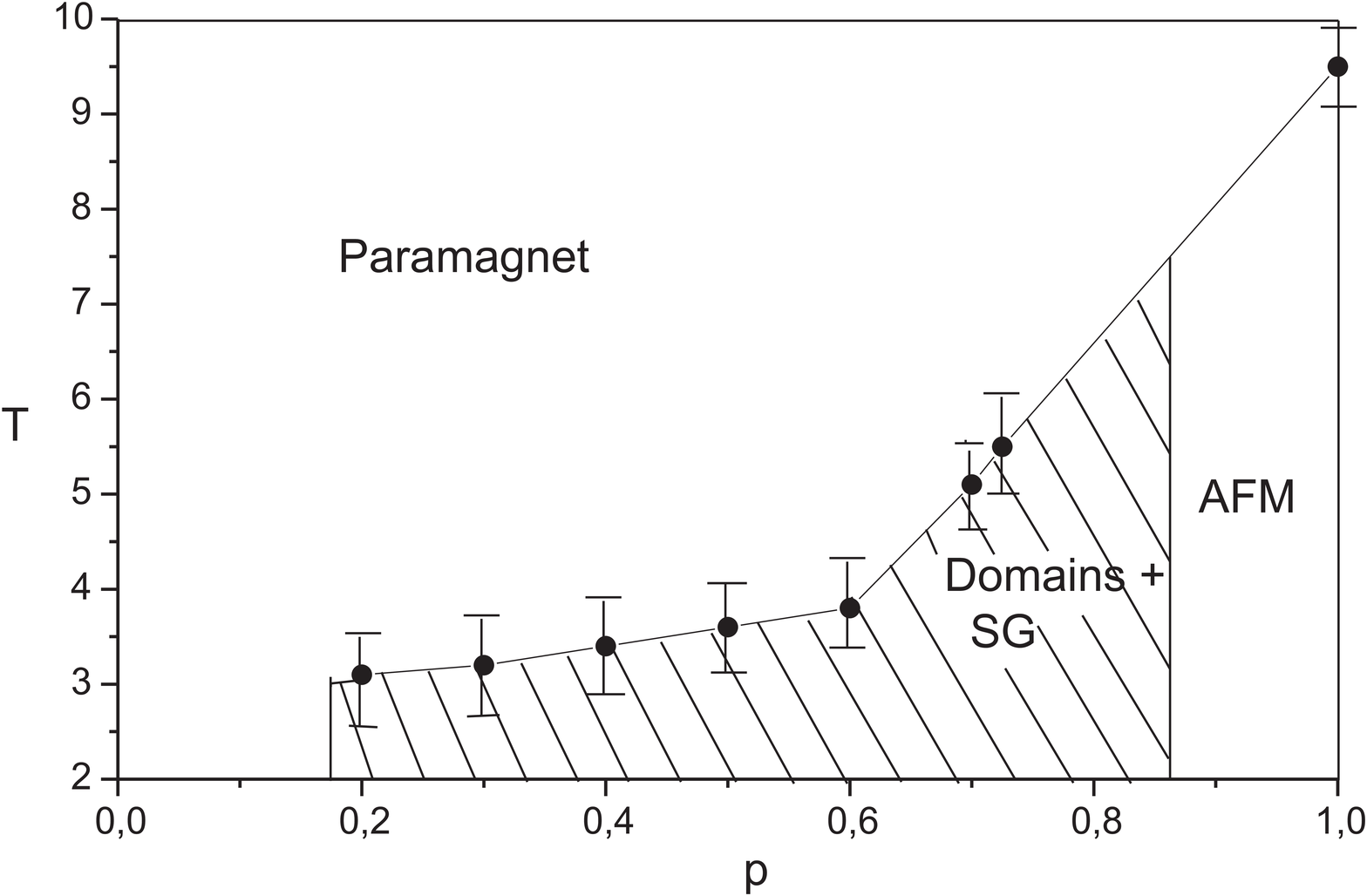} \\
\parbox[t]{0.47\textwidth}{\caption{ Temperature dependence of spin - glass order parameter: $h=3$; $L=24$;$p=0.2-0.7$, and $0.725$.}} \hfill
\parbox[t]{0.47\textwidth}{\caption{ Phase diagram for random-field Ising antiferromagnet at h = 3: PM = paramagnet; AFM = antiferromagnet; Domains+SG = domain structure and spin glass. }}
\end{figure}

We used the temperature and field dependence of magnetization, internal energy, and
specific heat to calculate the first-order phase transition lines. The $T-p$ phase
diagram shown in Fig. 5 summarizes the results obtained for $h=3$.

\section{ Conclusions }

The Monte Carlo simulations of thermodynamics of
the three-dimensional random-field Ising model performed
in this study demonstrate second-order phase
transition from paramagnetic to antiferromagnetic state
when the spin concentration is higher than pu and firstorder
phase transition from paramagnetic to mixed
phase consisting of antiferromagnetic and ferromagnetic
domains separated by spin-glass domains when
$p_c<p<p_u$, where $p_u$ and $p_c$ are vacancy and magnetic
percolation thresholds, respectively. When the spin
concentration is high, the system consists of antiferromagnetic
domains separated by spin-glass regions.
With decreasing spin concentration or increasing
applied magnetic field strength, both the number and
size of antiferromagnetic domains decrease, both the
number and size of ferromagnetic domains increase,
and the volume fraction of the spin-glass phase
decreases. It is shown that random magnetic field
induces a transition from antiferromagnetic to spinglass
ground state when $p_c<p<p_u$ in the three-dimensional
random-field Ising model with competing interactions
analyzed in this study.

\begin{center}
\bf Acknowledgmets
\end{center}

This work was supported by the Russian Foundation
for Basic Research (project nos.~04-02-17524 and
04-02-39000) and by the Ministry of Education of the
Russian Federation (grant no.~UR 01.01.230).


\begin{thebibliography}{99}

\bibitem{1}
A. B. Harris, J. Phys. C 7, 1671 (1993).
\bibitem{2}
R. Folk, Y. Holovatch, and T. Yavorski., Usp. Fiz. Nauk
173, 175 (2003) [Phys. Usp. 46, 169 (2003)].
\bibitem{3}
V. S. Dotsenko, Usp. Fiz. Nauk 165, 481 (1995) [Phys.
Usp. 38, 457 (1995)].
\bibitem{4}
Y. Imry and S.-K. Ma, Phys. Rev. Lett. 35, 1399 (1975).
\bibitem{5}
G. Parisi and N. Sourlas, Phys. Rev. Lett. 43, 744 (1979).
\bibitem{6}
J. Z. Imbrie, Phys. Rev. Lett. 53, 1747 (1984).
\bibitem{7}
J. Bricmont and A. Kupiainen, Phys. Rev. Lett. 59, 1829
(1987).
\bibitem{8}
A. P. Young and M. Nauenberg, Phys. Rev. Lett. 54, 2429
(1985).
\bibitem{9}
H. Rieger and A. P. Young, J. Phys. A 26, 5279 (1993).
\bibitem{10}
A. T. Ogielski and D. A. Huse, Phys. Rev. Lett. 56, 1298
(1986).
\bibitem{11}
 A. T. Ogielski, Phys. Rev. Lett. 57, 1251 (1986).
\bibitem{12}
J. Cardy, Phys. Rev. B 29, 505 (1984).
\bibitem{13}
D. P. Belanger and A. P. Young, J. Magn. Magn. Mater.
100, 272 (1991).
\bibitem{14}
G. S. Grest, C. M. Soukoulis, and K. Levin, Phys. Rev. B
33, 7659 (1986).
\bibitem{15}
F. Ye, L. Zhou, S. Larochelle, et al., Phys. Rev. Lett. 89,
157 202 (2002).
\bibitem{16}
D. P. Landau, Phys. Rev. Lett. 28, 449 (1972).
\bibitem{17}
H. Muller-Krumbhaar and D. P. Landau, Phys. Rev. B
14, 2014 (1976).
\bibitem{18}
V. V. Prudnikov, O. N. Markov, and E. V. Osintsev, Zh.
Eksp. Teor. Fiz. 116, 953 (1999) [JETP 89, 508 (1999)].
\bibitem{19}
K. Binder, Z. Phys. B 43, 119 (1981).
\bibitem{20}
V. S. Dotsenko, Usp. Fiz. Nauk 163 (6), 1 (1993) [Phys.
Usp. 36, 455 (1993)].
\bibitem{21}
E. Pytte, Y. Imry, and D. Mukamel, Phys. Rev. Lett. 46,
1173 (1981).

\end{thebibliography}
\end{document}